# Applicability of DFT + *U* to U metal and U-Zr alloy

Wei Xie,[1,] Chris A. Marianetti,[2] Dane Morgan[1,3,+]

[1]*Materials Science Program, University of Wisconsin-Madison, Madison, WI 53706, USA*
[2]*Department of Applied Physics and Applied Mathematics, Columbia University, New York, NY 10027, USA*
[3]*Department of Materials Science and Engineering, University of Wisconsin-Madison, Madison, WI 53706, USA*

[+]Corresponding author:
Dane Morgan
Email: ddmorgan@wisc.edu
Telephone: +1 608 265 5879
Address: 1509 University Avenue, Madison WI 53706, USA

# Abstract

In the Letter [Söderlind *et al*., J. Nucl. Mater. **444**, 356 (2014)] and Comment [Söderlind *et al*., Phys Rev B 90, 157101 (2014)], Söderlind *et al*. argue that 1) density functional theory (DFT) based on generalized gradient approximation (GGA) already models U metal and U-Zr alloy accurately, and 2) adding Hubbard *U* potential to DFT in DFT + *U* models U and U-Zr worse than DFT according to volume, bulk modulus, enthalpy and magnetic moments results they calculate or select from our recent study [Xie *et al*., Phys. Rev. B 88, 235128 (2013)]. They therefore conclude that DFT + *U* should be avoided for modeling U and U-Zr. Here we demonstrate in response to 1) that previously neglected and more recent experimental data indicate that DFT, even when implemented in all-electron methods, does not model the bulk modulus and elastic constants of αU very accurately. Furthermore, results from another all-electron study [Le Bihan *et al*., Phys. Rev. B 67, 134102 (2003)] suggest that Söderlind *et al*.'s claim that deficiency exists in our PAW calculation is unfounded and our results, including those that show DFT results compare unfavorably with experimental/computational references, are valid. We also demonstrate in response to 2) that Söderlind *et al*.'s arguments are unsound for three reasons. First, they focus on just the body centered cubic (BCC) phases γU and γ(U,Zr)—which at the *ab initio* modeling temperature of 0 K are thermodynamically and mechanically unstable and hence difficult to model and benchmark—as primary subjects of examination; however it is results for the stable phases mostly neglected by Söderlind *et al*. that are the primary evidence of our argument. Second, they make unfounded generalization of DFT + *U* results at selected $U_{\text{eff}}$ values to argue that DFT + *U* in general gives wrong or unsatisfactory results. Third and finally, some key points in Söderlind *et al*.'s criticisms of DFT + *U* are not well supported, including the claims that for γ(U,Zr) DFT + *U* at $U_{\text{eff}}$ = 1.24 eV gives positive deviations from linear dependence of composition that are unprecedentedly large and partially negative enthalpies of mixing that are inconsistent with the existence of miscibility gap in the experimental phase diagram. We therefore maintain our original conclusion that the accuracy of DFT has room for improvement and DFT + *U* can be of value for modeling U and U-Zr.

# 1. Introduction

In our recent two papers reporting CALPHAD (Xiong *et al*.[1]) and *ab initio* (Xie *et al*.[2]) modeling of the thermodynamics of U metal and U-Zr alloy, we argue that adding Hubbard $U$ potential to the standard functional of density functional theory (DFT)[3,4] in the so-called DFT + $U$ approach[5,6] could provide more accurate energetics for these systems compared to just DFT. Söderlind *et al*. object to our views first in a Letter[7] corresponding to our first article Xiong *et al*.[1] and later a Comment[8] to our second article Xie *et al*.[2]. Their objections consist of two main claims, which we label a) and b), on top of two and five arguments, respectively, which we label (a1)-(a2) and (b1)-(b5).

Söderlind *et al*.'s first main claim is that a) DFT already models ground state properties of the U and U-Zr systems sufficiently accurately that no beyond the DFT approach is necessary. To support this opinion, they cite about ten previous *ab initio* studies of the U and U-Zr systems that use only DFT. An exemplary one is an early study by Söderlind[9], which mainly concludes that (a1) DFT solved with the all-electron method full potential linearized muffin-tin orbital (FPLMTO) accurately reproduces the structural and elastic properties of the αU phase of elemental U metal. To refute our DFT results in Xie *et al*.[2] that contradict their claim in Ref.[9], they argue that (a2) our PAW calculations have "deficiency" and hence our obtained results are untrustworthy because the approach "seriously underestimates the volume of α-U" comparing to that from Söderlind's FPLMTO calculation[9].

Söderlind *et al*.'s second main claim is that b) adding Hubbard $U$ potential to DFT in DFT + $U$ deteriorates rather than improves DFT for modeling the U and U-Zr systems. To support this opinion, Söderlind *et al*. perform DFT + $U$ calculations at $U_{eff}$ = 2 eV on the γU phase of elemental U metal in their Letter[7] and argue that DFT + $U$ (b1) leads to "a grossly overestimated equilibrium volume", (b2) "substantially underestimates the bulk modulus", and (b3) predicts unphysical magnetic moments for γU. Later in their Comment[8], based on results they select from our DFT + $U$ calculations at $U_{eff}$ = 1.24 eV[2], Söderlind *et al*. refine criticism (b1) on volume by arguing that DFT + $U$ exacerbates the comparison to experiment for αU and "rather significantly overestimate volumes resulting in a strong deviation from Zen's law that is anomalous" with "the influence of SOC" on volume "greatly exaggerated" for γ(U,Zr). They also assert that (b4) DFT + $U$ calculation at $U_{eff}$ = 1.24 with spin-orbit coupling (SOC) included "gives negative enthalpies for a majority of the mixing", which they believe "is inconsistent with the known miscibility gap for the γ phase in the experimental phase diagram", and extend criticism (b3) on magnetic moments by stressing that DFT + $U$ calculations "predict significant spin and orbital magnetic moments for most phases of uranium and all γ-U-Zr alloys". Finally, besides model results, Söderlind *et al*. further argue that (b5) DFT + $U$ is "incomplete or inappropriate" because it needs the use of the model parameter $U_{eff}$ that they believe is "greatly different depending on the studied properties" for γ(U,Zr).

In this article, we will address the two major claims and the individual arguments of Söderlind *et al*. summarized in the previous two paragraphs that challenge the applicability of DFT + $U$ to U metal and U-Zr alloy. Before that, we clarify our views on two more minor issues Söderlind *et al*. bring up[7][8] that relate to opinions expressed in our work. The first issue is the nature of the strength of electron correlation in U and U-Zr. Söderlind *et al*. suggest that we view electron correlation in U and U-Zr to be strong, which they refute in their Letter[7]. We have not meant to suggest any such belief. This misunderstanding may be due to our use of the term "strong on-site correlation" once in Xiong *et al*.[1] (p 334) to refer to Hubbard $U$ potential in DFT + $U$, but may also be based on the idea that because we argue that DFT + $U$ can be of value for modeling U and U-Zr, we must also believe them to have strongly correlated electrons. We take this opportunity to clarify that the word "strong" is unwarranted when referring to Hubbard $U$ potential in DFT + $U$. We stress that DFT + $U$ can also be helpful for modeling materials that are not strongly correlated, for example FeAl[10]. In fact, based on the ratio of band width $W$ over Coulomb *U* from our calculations, we argue that U and U-Zr should be weakly to moderately correlated metals (p19 of Xie *et al*.[2]). The second minor issue is the importance of the relativistic effect of SOC in the U and U-Zr systems. In Xiong *et al*., we present *ab initio* enthalpy results from scalar relativistic and spin-orbit coupling included calculations, respectively (for convenience we have been referring them as noSOC and SOC calculations, respectively and we will follow this convention in this article) and also imply that including SOC can improve the accuracy of *ab initio* enthalpies. Söderlind *et al*. seem to interpret our doing so as equivalent to arguing that SOC is important and necessary to include in all *ab initio* calculations of U and U-Zr, which they also refute in their Letter[7]. Again, we do not mean to suggest that SOC is particularly important for the energetics of U and U-Zr. In fact, we have not given any explicit discussion of the importance of SOC for U and U-Zr in Xiong *et al*.[1] and in Xie *et al*.[2] we have argued that SOC affects ground state properties of U and U-Zr only marginally.

To avoid redundancy and potential confusion caused by material scattered in multiple articles in different journals, in this article, we reply in one place to objections to both of our articles[1,2] that Söderlind *et al*. make in both their Letter[7] and Comment[8]. We organize our rebuttals based on the two major claims a) and b) and their corresponding supporting arguments (a1)-(a2) and (b1)-(b5) that Söderlind *et al*. put forward. The body of this article next is divided into three sections. Sections 2 and 3 address Söderlind *et al*.'s arguments for a) and b), respectively. We believe many of Söderlind *et al*.'s criticisms of DFT + *U* are caused by confusion between different senses of "optimal" in our uses of the term "optimal $U_{\text{eff}}$" in various different contexts in Ref.[2], so in Section 3.1 we define "optimal $U_{\text{eff}}$" in both senses and clarify their values/ranges for the U and U-Zr systems. Finally, we summarize the main points of our reply in Section 4.

All our calculations in this article are done following the approaches described in detail in Xie *et al*.[2]. In particular, calculations are performed using the Vienna *Ab initio* Simulation Package (VASP)[11,12] based on the generalized gradient approximation (GGA) to the exchange-correlation potential parameterized by Perdew, Burke, and Ernzerhof[13]. The electron-ion interaction is described with the projected augmented wave (PAW) method[14] as implemented by Kresse and Joubert[12]. The simplified rotationally invariant form[6] of DFT + *U* that reproduces the standard DFT functional at $U_{\text{eff}}$ = 0 eV is used, which is the same form used in Söderlind *et al*.'s Letter[7]. As both DFT and DFT + *U* results referenced in this article are calculated using the GGA, this choice is not discussed further when comparing the two. Comparison based on the local density approximation (LDA) is not explored in our work because existing studies, for example in Refs.[15,16] consistently find that GGA treats U metal more satisfactorily than LDA.

# 2. Is there need of beyond the DFT approach for modeling U and U-Zr (rebuttal to argument (a))?

As mentioned in Section 1, Söderlind *et al*. put forward two arguments for their claim that a) DFT is already sufficiently accurate and no beyond the DFT approach is needed for modeling U and U-Zr. First, (a1) DFT calculations solved with the FPLMTO method performed by Söderlind[9] accurately reproduce the structural and elastic properties of αU phase of elemental U metal. Second, (a2) our PAW calculations that show DFT results for U and U-Zr[2] comparing unfavorably with experimental and/or computational references are invalid, because the method of PAW itself and/or our use of it have "deficiency", which Söderlind *et al*. support by comparing the volume of αU between PAW calculations of us and FPLMTO of Söderlind[9]. We disagree with Söderlind *et al*.'s claim a) and in this section we refute their supporting arguments (a1) and (a2) in sequence.

## 2.1. Is DFT solved with all-electron methods sufficiently accurate (rebuttal to supporting argument (a1))?

We show that Söderlind *et al.*'s argument (a1) is spurious because issues exist in the experimental data against which Söderlind[9] benchmarked his theoretical values. Based on surveys of experimental data that include those neglected by Söderlind[9] and more recent experiment, this section demonstrates that DFT, even solved with all-electron methods like FPLMTO[9] still models the bulk modulus and elastic constants of αU unsatisfactorily, which motivates pursuit of beyond the DFT approaches like DFT + *U*.

### 2.1.1. Bulk modulus of αU

Until recently the belief that DFT models the bulk modulus of αU very accurately seems to be widely held and rarely disputed by the computational community. For example, Söderlind[9] reports that DFT solved by FPLMTO reproduces the bulk modulus of αU measured with X-ray by Yoo *et al.*[17] almost perfectly. To our knowledge, Le Bihan *et al.* [18] are among the few to challenge this view by suggesting that an issue exists in the early X-Ray experiments [17,19-21] used to benchmark theory. Despite having been neglected by most later studies, which continue to benchmark theoretical results against those early X-Ray results[17,19-21], Le Bihan *et al.*[18] have made a solid case based on data from a number of studies using different experimental techniques in addition to their own X-ray study. Below we elaborate Le Bihan *et al.* [18]'s argument and cite more recent X-ray results to show that DFT models the bulk modulus of αU unsatisfactorily.

As shown in Figure 1, results from early X-ray studies[17,19-21] published before 2000, which range between 125-157 GPa at room temperature (abbreviated as r.t. hereafter), are considerably higher than those from all the other experimental techniques, which are 114.7 GPa[22] and 114 GPa[23] at r.t. from ultrasound, 112 GPa[24] at r.t. from neutron diffraction, 101 GPa[25] at r.t. from mechanical displacement and 107 GPa[26] at ~0 K from specific heat. According to Le Bihan *et al.* [18], the issue is that those early X-ray values[17,19-21] are overestimated substantially due to nonhydrostatic stress effects. Using a quasihydrostatic media nitrogen Le Bihan *et al.* obtain a value of 104 GPa[18], which is much closer to values from the other techniques. The controversy should now be considered resolved thanks to a very recent X-ray study by Dewaele *et al.*[27] that obtains a value of 114.5 GPa at 298 K using a more quasihydrostatic media helium and properly calibrated pressure gauge. Therefore, bulk modulus for αU at 298 K is expected to be in the range 101-115 GPa with 114.5 GPa being a currently best estimate. The value seems to be smaller at lower temperature, as suggested by the specific heat[26] result of 107 GPa at ~0 K, but that value may be affected by charge density waves at and below 43 K [28] (charge density wave is not modeled in the *ab initio* calculations at 0 K referenced in this article). A more meaningful reference is for example 111.5 GPa calculated from the elastic constants measured at 46 K by Fisher and McSkimin[29] with ultrasound technique. Lawson and Ledbetter[30] also obtain an estimated 0 K value of 114.4 GPa based on the ultrasound data between 298 and 923 K that Fisher compiles in Ref.[31]

In comparison, Figure 1 shows that values ranging from 133 to 148 GPa are obtained in DFT calculations with SOC included using FPLMTO[9,18] or full potential augmented plane waves (FPLAPW) [32,33], which are larger than the 46 K experimental reference value of 111.5 GPa by 21.5-36.5 GPa (19-33%). Moreover, according to Söderlind[9], neglecting SOC leads to an increase in bulk modulus by about 8% for αU, so the error increases to approximately 27-41% if calculated without including SOC. Such error magnitudes are outside the range of what is expected from DFT. To put them in context, for the ground state phase of most transition metals the corresponding DFT errors for bulk modulus are 10% or less according to Refs.[34,35], even though the calculations are done therein using PAW, not an exact all-electron method. This analysis suggests that the bulk modulus of U is not well predicted by standard DFT even with all-electron methods. We will revisit bulk modulus below in Section 3.2.2 and show that DFT + U can improve DFT on modeling the bulk modulus of both αU and γU, in agreement with the case for enthalpy and volume that we make in Ref.[2].

### 2.1.2. Elastic constants of αU

We discuss in this section the case for elastic constants of αU ($c_{ii}$ and $c_{ij}$, where $ii$=11 to 66 and $ij$=12, 13, 23), the data of which are given in Table 1. On the experiment side, Fisher and coworkers have measured all the 9 independent elastic constants of αU with ultrasound technique first at 298 K[22], then from 78 to 298 K[36], from 43 to 78 K[29], from 298 to 923 K[31], and finally below 43 K[37]. On the theory side, we only find one[9] DFT calculation of the elastic constants of αU (SOC is included) that employs all-electron method, although several studies employing methods that are not all-electron are available[27,38-40]. To make the comparison more accessible, we will use the all-electron results of Söderlind[9] to represent typical DFT values. In that study, Söderlind[9] compares DFT values for both $c_{ii}$ and $c_{ij}$ against Fisher and McSkimin's ultrasound data at 298 K[22] and for only $c_{ii}$ against the values that he linearly extrapolates to 0 K, and concludes that "for the most part elastic constants agree well with experiment". Below we perform an independent analysis of both Fisher and coworkers' experimental and Söderlind's computational data and show that they do not support Söderlind's conclusion based on the comparisons at either 298 K or 0 K.

At 298 K, Söderlind states[9] that his FPLMTO results overestimate the 298 K experimental data of Fisher and McSkimin[22] "with an average amount of about 27%". According to our calculations, the mean absolute error (MAE) and root mean square error (RMSE) (see Table 1 for definitions) for Söderlind's[9] DFT values referencing to Fisher and McSkimin's[22] ultrasonic values at 298 K are 30% and 38%, respectively.

At 0 K, Söderlind[9] estimates $c_{ii}$ using linear extrapolation based on the values and slopes of the experimental data at 100 K from the article of Fisher and McSkimin[29], but does not estimate $c_{ij}$ because he finds that "no such temperature variation has been measured" for $c_{ij}$. We believe this 0 K estimation can be improved on two aspects. On the one hand, $c_{ij}$ could be included because it turns out that the temperature variation of $c_{ij}$ has been measured. Fisher and McSkimin have measured and plotted in Ref.[29] and Fisher has tabulated in Ref.[31] all the 9 elastic constants of both $c_{ii}$ and $c_{ij}$ measured between 44 and 923 K. Both these references are available at the time of Söderlind's estimation[9] but perhaps these data are missed. In addition, we believe Söderlind's effective 0 K values for $c_{ii}$ obtained using the slopes at 100 K for linear extrapolations to 0 K are probably overestimated, because Fisher and McSkimin[29] show that for $c_{11}$ the already positive slope at 100 K becomes more positive around 50 K while for $c_{22}$ to $c_{66}$ the sign of their slopes change from negative at 100 K to positive around 50 K. These changes in slopes are continuous and thus are intrinsic for perfect αU bulk that should be included when estimating the temperature effects, unlike the discontinues changes around 43 K due to charge density wave.

As alternatives to Söderlind's linearly extrapolated 0 K values for only the 6 $c_{ii}$'s, we provide in Table 1 three different sets of low temperature experimental references for all the 9 elastic constants of both $c_{ii}$ and $c_{ij}$. The first set contains values extrapolated to 0 K by McSkimin and Fisher from polynomial fitting of the their data measured between 78 and 298 K[36]. Comparing to Söderlind[9]'s, McSkimin and Fisher's extrapolation is not limited to linear functional form, considers values and curvatures of experimental data at multiple temperatures, and therefore is expected to be a better fit. However, it still does not consider those data near 50 K where important changes of slopes happen. Such effects should have been included in the second set of data, which are measured by Fisher and McSkimin[29] at 46 K. This temperature is close to 43 K at and below which charge density waves onset while smaller than 50 K about which changes of slopes happen, so we believe the data at 46 K are of particular value for extrapolation to 0 K. Finally, Table 1 also lists the data measured at 4.2 K, the lowest temperature reached in the experiment by Fisher and Dever[37]. Although closest to 0 K, they are affected by charge density waves, which reduces their value as reference. After describing these issues of experimental reference, we are ready for a comparison. Table 1 shows that MAE for Söderlind's DFT data are 30%, 34%, 44% and 29%, respectively when referencing to the four sets of data at 298, 46, 4.2 and 0 K. The corresponding RMSE are 38%, 49%, 67% and 37%, respectively. It is expected that neglecting SOC will further increase the error. The magnitudes of the error, even on the lower end, are again fairly large, consistent with the case for bulk modulus discussed above. Again to put them in context, αU's RMSE error of 49% (referenced to the 46 K data) or 37% (referenced to the extrapolated 0 K data) from Söderlind's DFT calculations using FPLMTO is significantly larger than the RMSE errors of < 20% for most transition metals found in Shang *et al.*'s DFT calculations using PAW[34]. This analysis suggests that the $c_{ii}$ and $c_{ij}$ elastic constants of U are not well predicted by standard DFT.

## 2.2. Is there a significant error associated with using PAW vs. all-electron methods for αU (rebuttal to supporting argument (a2))?

Söderlind *et al.* argue (a2) that our DFT calculation[2] using PAW "seriously underestimates the volume of α-U" (20.07/20.06 Å$^3$/atom) when comparing to the noSOC/SOC values of 20.67/20.40 Å$^3$/atom from Söderlind's[9]'s FPLMTO calculations, and the difference of 0.60/0.34 Å$^3$/atom (about 3%/2%) proves the existence of a "deficiency" in our PAW calculations. We show that the difference between our PAW[2] and Söderlind's FPLMTO[9] can be caused by different GGA functionals, structural relaxations, and implementations of SOC used between the two sets of calculations in addition to the PAW approximation, and that this discrepancy therefore cannot be taken to point to a deficiency in our PAW calculations.

As we have already cited in our discussion of the difference in our previous article[2], another FPLMTO study of αU by Le Bihan *et al.*[18] obtains a volume value of 20.34 Å$^3$/atom calculated with SOC included, differing from Söderlind's SOC value by 0.33 Å$^3$/atom (2%). The fact that even using the same FPLMTO method still produces a difference accounting for a majority of the difference between the two different methods of PAW and FPLMTO should have persuaded most rational observers that the latter differences are probably within reason. However, to clear any doubt we now analyze quantitatively the origins of the differences. Besides (1) the PAW approximation (Xie *et al.*[2]) vs. all-electron (Söderlind[9] and Le Bihan *et al.*[18]), we believe two computational details are also likely sources of difference : (2) GGA functional—PW-91 (Söderlind[9]) vs. PBE (Xie *et al.*[2] and Le Bihan *et al.*[18]), and (3) structural relaxation—sequential iterative relaxation (Söderlind[9]) vs. parallel relaxation with conjugate gradient algorithm (Xie *et al.*[2] and Le Bihan *et al.*[18]) of the lattice constants and atomic positions. Our study is the same as Le Bihan *et al.*'s on both (2) and (3), so we can to a good approximation estimate source (1) to cause a difference of about 20.34 (Le Bihan *et al.*)–20.07 (Xie *et al.*) = 0.27 Å$^3$/atom. Effect 2) can be estimated to cause a difference of about 20.23–20.07=0.16 Å$^3$/atom, where the latter value of 20.23 Å$^3$/atom is from another DFT SOC calculation of us as in Ref.[2] except using PW-91. Based on them, effect 3) can be estimated to be about 20.67 (Söderlind)–20.34 (Li Bihan *et al.*)–0.16 (effect (2))=0.17 Å$^3$/atom. The above estimations are based on results calculated with SOC. To analyze the noSOC case, since no corresponding value is reported by Le Bihan *et al.* we have to take an indirect approach, namely by assuming that sources (2) and (3) estimated above to be unchanged because they should be relatively independent of SOC. As given above, the difference between our PAW and Söderlind's FPLMTO values calculated without SOC included is 0.34 Å$^3$/atom. Subtracting those due to effects (2) and (3) we can estimate that without SOC the PAW approximation only amounts to a difference of 0.01 Å$^3$/atom, so even our previous estimation of 0.27 Å$^3$/atom based on the data calculated with SOC included may have significantly exaggerated source (1) because it may have included the difference due to source (4) different implementations of SOC between FPLMTO and PAW. In short, the discrepancies between our PAW and the particular FPLMTO study by Söderlind[9] are within the range of what can be caused by different numerical and methodological treatments, and declaring the difference proof of a "deficiency in the PAW treatment for uranium" of us is unreasonable.

# 3. Can DFT + $U$ improve DFT on modeling U and U-Zr (rebuttal to argument (b))?

In this section, we reply to Söderlind *et al.*'s criticisms of applying DFT + $U$ on modeling U and U-Zr (i.e., claim b) and its supporting arguments (b1)-(b5)). Most of the criticisms of Söderlind *et al.* are based on results for the high temperature body centered cubic (BCC) phases γU and γ(U,Zr). We believe focusing on γU and γ(U,Zr) introduces unnecessary complications into the debate because they are both thermodynamically and mechanically unstable at 0 K[2,41]. Both types of instability mean that very limited low temperature data are available to benchmark *ab initio* methods employed in our calculations at 0 K—more specifically, direct experimental data are limited and CALPHAD results are controversial. Moreover, the mechanical instability makes it very difficult to accurately model these phases *ab initio* at 0 K. In order to better ground the discussion we will therefore include relevant results for well-characterized stable phases, in particular the ground state phase αU of U metal, as well as make every effort to identify and extrapolate the best low temperature reference data for γU and γ(U,Zr). In the following we first clarify the issue of "Optimal $U_{eff}$" in Section 3.1 and then reply to Söderlind *et al.*'s arguments (b1) to (b5) in Sec. 3.2.

## 3.1. "Optimal $U_{eff}$" for U atoms in the U and U-Zr systems

Result of DFT + $U$ depends critically on the model parameter $U_{eff} = U - J$ in the implementation of DFT + $U$[6] that we have been using. While in their Letter[7] Söderlind *et al.*'s disagreement with our view on the applicability of DFT + $U$ for modeling U and U-Zr may be caused in part by their choice of $U_{eff} = 2$ eV in their evaluations of DFT + $U$ based on γU, a value that we believe is too large according to our fitting and not one we propose using, we believe additional confusion is caused by Söderlind *et al.*'s interpretation of the "Optimal $U_{eff}$" that we have proposed based on our validations[2] of DFT + $U$. There are two subtly different types of optimal $U_{eff}$ in our work, and we have perhaps not distinguished them as clearly as we should have. It seems that the two different types of optimal $U_{eff}$ are somewhat mixed together in Söderlind *et al.*'s Comment[8]. Therefore, before we delve into Söderlind *et al.*'s specific criticisms, in this section we first define explicit terminologies for each type of "optimal $U_{eff}$" and clarify their values/ranges.

The first type of optimal $U_{eff}$ that we define is the "single-structure optimized $U_{eff}$" (previously called "empirical $U_{eff}$" in Ref.[2]), which is the $U_{eff}$ with which DFT + $U$ best reproduces one or more physical properties (e.g., enthalpy) of a single material (i.e., crystal structure). If there are multiple properties involved, then the single-structure optimized $U_{eff}$ will be the value that best represents all the data available, which quantitatively can be determined through least-square fitting. In reality, however $U_{eff}$ for different properties of a same structure is quite close, if not exactly the same, as we will discuss in Section 3.2.5 in reply to Söderlind *et al.*'s argument (b5). Moreover, in Ref.[2], we find composition affects negligibly Hubbard $U$ of U and U-Zr that we estimate theoretically using the linear response approach[42], so approximately each *single* phase (i.e., parent crystal lattice) has a "single-structure optimized $U_{eff}$". As expected, single-structure optimized $U_{eff}$ values vary for different phases, and for the U and U-Zr systems that contain multiple phases, single-structure optimized $U_{eff}$'s for individual phases are distributed within a range. The second type of optimal $U_{eff}$ that we define is the "multi-structure optimized $U_{eff}$" (previously called "statistical optimal $U_{eff}$" in Ref.[2]), which is the $U_{eff}$ with which DFT + $U$ best reproduces (with least-squares fitting) all the values of one or more physical observables of *multiple* phases together. We use the term multi-structure optimized $U_{eff}$ to emphasize that this value is optimal only when considering the average of multiple phases but may or may not be optimal for an given phase. In Ref.[2] we determine all values for single-structure and multi-structure optimized $U_{eff}$ by fitting to the enthalpy for different phases; other properties like volume are also compared but not quantitatively fitted in Ref.[2].

Based on our results in Ref.[2] and additional results to be presented below, we determine that single-structure optimized $U_{eff}$ varies approximately in the range of 1- 1.5 eV for the 8 solid phases of U and U-Zr explored in Ref.[2] while the overall multi-structure optimized $U_{eff}$ for the U and U-Zr systems is 1.24 eV. For example, γU's single-structure optimized $U_{eff}$ should be around 1 eV because both volume in Ref.[2] and bulk modulus, to be shown below, reproduce the corresponding experimental references near 1 eV. Note we have been using $J = 0.51$ and the somewhat awkward values of 0.99 and 1.49 eV (from $U - J = 1.50-0.51$ and 2.00-0.51, respectively) are the exact values of $U_{eff}$'s that are actually used, but more often their rounded values of 1 and 1.5 eV are used in our discussion. Extending fitting from data on γU to include all of the BCC phase solid solution γ(U,Zr) still produces a single-structure optimized $U_{eff}$ of approximately 1 eV based on enthalpy and volume comparisons. We will discuss this value in more detail below. Different from γU and γ(U,Zr), αU's single-structure optimized $U_{eff}$ may be as large as 1.5 eV, and so is the value for δ(U,Zr). In general, to best describe an individual phase, its own "single-structure optimized $U_{eff}$" should be used, and that is why we use $U_{eff} = 1$ eV for γ(U,Zr) and $U_{eff} = 1.5$ eV for δ(U,Zr) in Ref.[1]. We unfortunately have neglected to report the two $U_{eff}$ values used in Ref.[1]. This is an oversight we are glad to be able to correct here and it may have led to some confusion on the issue of the accuracy of DFT + $U$.

These optimal $U_{eff}$ values that we establish from empirical fitting are also found to be reasonable based upon theoretical estimation of the Hubbard $U$ for U atoms in the U and U-Zr systems. Using the linear response approach[42], we estimate that U and U-Zr's Hubbard $U$ is in the range of 1.9 – 2.3 eV. Another calculation also obtains a value of about 2 eV for U metal[43]. While these theoretical values are somewhat larger than our empirically determined range of single-structure optimized $U_{eff}$ of 1-1.5 eV, this discrepancy is to be expected, because DFT + $U$ is the Hartree-Fock approximation to DFT + DMFT (dynamical mean field theory) that is expected to overestimate the effects of the Hubbard $U$[44]. Therefore, it is natural that one would arrive at somewhat smaller values of the Hubbard $U$ when fitting DFT + $U$ results to observables like enthalpy as compared to direct computation of $U$ theoretically using approaches like the linear response.

## 3.2. Addressing Söderlind *et al.*'s criticisms of DFT + *U* (rebuttal to arguments (b1)-(b5))

In this section we address Söderlind *et al.*'s criticisms of DFT + $U$, which are made primarily on the high temperature BCC phases γU and γ(U,Zr). Söderlind *et al.* argue that DFT + $U$ gives inaccurate values for b1) volume, b2) bulk modulus, b3) enthalpy, and b4) magnetic moments. They also argue that b5) the value of optimal $U_{eff}$ for γ(U,Zr) is "greatly different depending on the studied properties".

### 3.2.1. Volume (rebuttal to argument (b1))
Söderlind *et al.* argue that (b1) DFT + $U$ gives worse volume than DFT because for αU it exacerbates the error vs. experiment and for γ(U,Zr) it gives volume values that they believe are too large in terms of absolute value, show excessive deviation from linear volume-composition dependence, and have too much expansion due to SOC. We address these criticisms sequentially next.

*αU*

Söderlind *et al.* support (b1) in the Comment[8] by arguing that for the volume of αU our DFT + $U$ ($U_{eff}$=1.24 eV) values of 20.75/20.94 Å$^3$/atom calculated using PAW without/with SOC included agree less well with the experimental reference of 20.53 Å$^3$/atom than the corresponding DFT values of 20.40/ 20.67Å$^3$/atom from Söderlind's FPLMTO calculations[9]. As we have addressed in our original article[2], the comparison is not valid to support Söderlind *et al.*'s conclusion because it involves other factors that are entangled with the difference between DFT and DFT + *U*.

As we have analyzed above in Section 2.2, calculations using different methods or even the same method but in different studies can give different results due to differences in subtle numerical details that would mask the relative change from DFT to DFT + *U*. To assess the relative accuracy of DFT and DFT + *U* it is best to keep all other numerical details the same, as we do in Ref [2]. If this method is FPLMTO and the DFT result is Söderlind's 20.67/20.40 Å$^3$/atom, then because DFT + *U* predicts larger volume than DFT, it indeeds gives worse comparison. However, we have estimated above in Sec. 2.2 that Söderlind's DFT values may have been overestimated due to structural relaxation approximately by 0.17 Å$^3$/atom, and if that effect is included then DFT + *U* actually can still improve the volume of αU. The case is even more strongly supported if one uses the DFT SOC value of 20.34 Å$^3$/atom from Le Bihan *et al.*[18]. We note that all these differences between different methods and between DFT and DFT + *U* are relatively small (1-3%) and our view[2] has been that the case for this specific system of αU is debatable[2]. A more robust comparison would need to examine more systems. In Ref.[2], we have checked all the solid phases of U and U-Zr stable in the current accepted ambient pressure phase diagrams using the same PAW method. We believe the conclusion that DFT + *U* can in general improve the volume of U and U-Zr[2] reached from such comparisons is more robust than one that is based only on αU.

*γ(U,Zr)*

Söderlind *et al.*'s main support for (b1) is from the results for γ(U,Zr) (including γU implicitly as an end member of γ(U,Zr)). First, they argue that their DFT + *U* calculations[7] at $U_{eff}$ = 2 eV give too large volume for γU when compared with some experimental values compiled by Donohue[45]. Moreover, they argue[8] that our DFT + *U* calculations[2] at $U_{eff}$ = 1.24 eV "rather significantly overestimate volumes resulting in a strong deviation from Zen's law that is anomalous" and "the influence of SOC" on volume is "greatly exaggerated" for γ(U,Zr). Our view for γU/γ(U,Zr), which has been stated in our original articles[1] and[2] has always been that significant uncertainty exists due to their mechanical instability and lack of low temperature data and a consensus may not be possible at present. However, for the purpose of replying to Söderlind *et al.*'s criticisms, here we make our best attempt at a comparison. Below we perform a thorough assessment of the currently available experimental data for the volume of γ(U,Zr) and compare *ab initio* data from our own PAW calculations as well as all-electron ones from the literature to the assessed most reliable experimental data. Doing so, we find that the currently available data are biased towards supporting that 1) DFT underestimates the volume of γ(U,Zr); 2) DFT + *U* at $U_{eff}$ = 1.24 eV— the multi-structure optimized $U_{eff}$ for overall U and U-Zr does give large error of volume for γ(U,Zr) on the U-rich end, but 3) DFT + *U* at $U_{eff}$ = 1 eV, the single-structure optimized $U_{eff}$ for γ(U,Zr), improves the volume for γ(U,Zr) compared to DFT. These results contradict Söderlind *et al.*'s general proposition that DFT + *U* exacerbates the volume of γ(U,Zr) compared to just DFT.

We first review existing experimental data, as given in Table 1 a) and plotted in Figure 2. We have done such a review already for the two end members γU and βZr in Ref.[2] and thus for each of them only the experimental values that are evaluated to be most accurate will be referenced here—Lawson *et al.*[46] for γU and Heiming *et al.*[47] for βZr. Both these experimental values are measured at high temperature and are corrected to give approximate 0 K volumes appropriate for comparison to *ab initio* values (see the supplementary materials of Ref.[2] for the review of experimental data for γU and βZr and the details of the temperature correction). For γ(U,Zr) at intermediate compositions, we find three experimental volume measurements—Huber and Ansari[48], Akabori *et al.*[49] and Basak *et al.*[50]. Both Huber and Ansari and Basak *et al.* measure quenched samples at room temperature. Basak *et al.* [50] find that two of the three quenched γ(U,Zr) samples actually have phase separated into mixtures of γ(U,Zr) and δ(U,Zr), and thus we will only refer to the value from the sample that do not phase separate next. We suspect phase separation may have happened in some of Huber and Ansari's samples as well, because their data show a convex shape vs. composition. Such a shape differs from all other experimental and *ab initio* data and is counterintuitive for a phase separating alloy like γ(U,Zr), and thus is possibly incorrect. In particular, we expect the problem to be more severe near δ(U,Zr)'s stable composition range of about 60-80 at.%Zr. Therefore, we should put less weight on the data from Huber and Ansari, despite the fact that it is the only study we find that has measured a wide composition range. The only data point measured directly at high-temperature for γ(U,Zr) is from Akabori *et al.*[49], which is particularly valuable as it probes the true structure of γ(U,Zr) that is only stable at high temperatures. The original value again has been corrected to give an approximate 0 K volume appropriate for comparison to *ab initio* values, as done above for the other high-temperature data. As we can see in Figure 2, Basak *et al.*'s and Akabori *et al.*'s values are very close, so they should be considered especially trustworthy.

Next we compare our *ab initio* results calculated using PAW to other *ab initio* studies of γ(U,Zr) in the literature to clarify any issue on differences between *ab initio* methods. For this purpose, let us look at Table 1 b) and Table 1 c). They show that our DFT noSOC/SOC calculations using PAW predict γU's volume to be 0.2/0.4 Å$^3$/atom smaller than Söderlind *et al.*[7]'s calculated using FPLAPW (magenta square symbols). For βZr, our DFT calculations using PAW obtain essentially the same result as Landa *et al.*[51]'s DFT calculations using FPLMTO (open orange square symbol; estimated from Fig 9 of Ref. [51]). These results show that our noSOC/SOC PAW results are about 1%/2% smaller than the FPLAPW calculations for γU, which is consistent with the situation for αU that we have analyzed in Section 2.2 and should be likely in part due to the PAW approximation but also come from other differences in the two calculations. Additional calculations from Landa *et al.*[51] (brown open symbol and dash line) using the Korringa-Kohn-Rostoker method in the Atomic Sphere Approximation (KKR-ASA) obtain values that are larger than all the other three calculations using FPLAPW, FPLMTO and PAW by approximately 1.0 Å$^3$/atom (5%) in the whole composition range, even for βZr. It should be noted that KKR-ASA values are from model temperature of 300 K and the other three calculations from 0 K, but the temperature effect should be quite small (<0.2 Å$^3$/atom based on our estimation in Ref.[2]) and cannot explain the majority of the large discrepancy of about 1.0 Å$^3$/atom. This comparison suggests that the range of DFT values can be relatively large depending on the methods used. However, if we exclude the KKR-ASA results as involving additional approximations, then the discrepancies between PAW and FPLAPW for γU are still significantly smaller than the difference between the *ab initio* DFT results and the experiment values, which are about 6% for PAW and 5% for FPLAPW. This comparison suggests that in exploring DFT + $U$ effects on volume our PAW calculations can be considered to yield results that are close enough compared to the best *ab initio* calculations using all-electron methods like FPLAPW to allow meaningful comparison to experiments and assessment of the effects of adding Hubbard $U$ potential.

Before we compare *ab initio* results to experimental data for γ(U,Zr), we stress again that such a comparison will face major uncertainty and is potentially very misleading. We feel this way for three reasons. Firstly, γU and γ(U,Zr) are high temperature phases and thermodynamically unstable at 0 K, so no corresponding experimental volume result at low temperature is directly available. Some data measured at room temperature on quenched samples are available, but their reliability is often uncertain. As a result, most of the time we have to extrapolate experimental data from the actual measurement temperatures over 1000 K to 0 K, which necessarily introduces considerable uncertainty. Secondly and more importantly, γU and γ(U,Zr) at many compositions are also mechanically unstable[2,41] and can only be modeled in 0 K *ab initio* calculations simply with lattice shape and ion position constrained relaxations (other approaches could be used (e.g., Refs. [52,53]), but are outside the scope of our present work). Thirdly, for reasons we do not full understand at this time, γU and γ(U,Zr) seem particularly susceptible to the problem of metastable solutions with DFT + $U$[54], which otherwise can be quite well mitigated with approaches like U-ramping[55], which we have applied in Ref.[2]. For example, we recently find that for γU calculated by DFT + $U$ at $U_{eff}$ = 1 eV, another solution exists that is 0.001 eV/atom lower in energy than the solution we have reported in FIG. 6 of Ref. [2]. While this energy difference is not significant and does not impact any conclusion of Ref.[2], this new solution has a volume of 21.18 Å$^3$/atom, while the old solution, despite being very close in energy, has a volume of 21.51 Å$^3$/atom.

However, to address the criticisms of Söderlind *et al*. about the calculated γU's volume, we still proceed and compare our *ab initio* volume data for γ(U,Zr) (including the two end members γU and βZr), calculated using PAW by us or all-electron methods from the literature, to the above reviewed experimental results in Figure 2. The numerical values are also given in Table 1. We make this comparison on three aspects.

In the first aspect of our analysis of the volume of γ(U,Zr) we focus on the absolute values of volume in Fig. 1 a). Söderlind*et al*. argue that their DFT + *U* calculations[7] at $U_{eff}$ = 2 eV gives too large volume for γU, while our DFT + *U* calculations[7] at $U_{eff}$ = 1.24 eV overestimates the volumes of γ(U,Zr) significantly. We agree with these statements but show that DFT + *U* at $U_{eff}$=1 eV gives improved volume compared to DFT for γ(U,Zr). Fig. 1 a) shows that our DFT calculations using PAW underestimate the volume of γ(U,Zr) in the whole composition range. The error is larger at larger U concentration. For example, when comparing to the estimated 0 K experimental volumes, for the γU end member, the error is about 0.8 Å$^3$/atom from both noSOC and SOC calculations of us using PAW, and about 0.7/0.4 Å$^3$/atom from noSOC/SOC calculations of Söderlind *et al*.[7] using FPLAPW. In comparison, DFT + *U* gives larger volume than DFT, and at $U_{eff}$ = 1, our DFT + *U* volume results are closer to the most reliable values discussed above than DFT (specifically Lawson *et al*.[46] for γU (filled black pentagon symbol), the high-temperature γ(U0.707Zr0.293) data from Akabori *et al*.[49] (filled black triangle symbol), and the quenched γ(U0.723Zr0.277) sample of Basak *et al*.[50] that does not phase separate (filled black rhombus symbol)). The improvement is particularly significant for γU. However, at a slightly larger $U_{eff}$ of 1.24 eV, DFT + *U*, especially when SOC is included, gives volumes that are larger than the experimental values on the U-rich end with the error being most significant for γU. The exact source of the error at $U_{eff}$ = 1.24 eV is still unknown, but the large changes in predicted volume suggest that something significant has changed in the electronic structure. Indeed, we can see in FIG.10 of Ref.[2] that for γU and γ(U,Zr) the density of states (DOS) differs significantly between those calculated by DFT and by DFT + *U* at $U_{eff}$ = 1.24 eV, while for other systems like αU, βU and δ(U,Zr) the DOS is still similar between DFT and DFT + *U* at $U_{eff}$ = 1.24 eV.

In the second aspect of our analysis of the volume of γ(U,Zr), we consider how much volume deviates from linear dependence of composition. Söderlind, *et al*. argue[8] that our DFT + *U* calculations at $U_{eff}$ = 1.24 eV with SOC included gives volume of γ(U,Zr) that deviates from linear composition dependence (Zen's law[56]) to an extent that they think is "unprecedented". We disagree with this assertion because it is based only on false visual impression given in FIG. 1 of Söderlind*et al*.'s Comment[8]. Specifically, their volume axis spans a very small scale of about 0.60 Å$^3$/atom (3%) thanks to the fact that values for the two end members γU and βZr are close. As a result, although quantitatively the largest deviation from linearity is only 0.49 Å$^3$/atom at 50 at.%Zr from our DFT + *U* (1.24 eV) SOC calculation, it accounts for 0.49/0.60*100%=82% of the axis scale, giving the impression that it is larger than might be expected. In reality however, the deviation from our DFT + *U* (1.24 eV) SOC calculation is even slightly smaller than the corresponding maximum value from our DFT-SOC calculation, 0.52 Å$^3$/atom at 50 at.%Zr. Quantitatively, deviation from linear volume-composition relationship can be characterized with "volume of mixing" defined as $V^{mix}_{\gamma(U,Zr)} = V_{\gamma(U,Zr)} - (1-x)V_{\gamma U} - xV_{\beta Zr}$, where *x* is Zr mole fraction, which is plotted in Fig. 1 b. This figure shows that at $U_{eff}$ = 1 eV DFT + *U* gives larger volume of mixing than DFT, reaching a maximum of about 1 Å$^3$/atom (5%) at 0.25 mole fraction of Zr, although at $U_{eff}$ = 1.24 eV with SOC included it actually drops back to be very close to DFT's. As reference, we calculate the volume of mixing for the two experiment data of Akabori *et al*.[49] and Basak *et al*.[50] by using the end member volumes from Lawson *et al.*[46] (for γU) and Heiming *et al.*[47] (for βZr). Fig. 1 c). shows that DFT gives volume of mixing that is closer to the two particular estimated experimental points than DFT + *U* at $U_{eff}$ = 1 eV by about 0.2 Å$^3$/atom (1%). However, given that the alloy and the end member data used to calculate the two reference values come from different experimental sources, the experimental volume data must be extrapolated from high temperature to 0 K (which is an ~0.5 Å$^3$/atom effect for γU), and that we are modeling a dynamically unstable phase with constrained *ab initio* calculations, we do not think that such a scale of difference is very meaningful. Furthermore, we point out that significant deviation from Zen's law is not in itself a sign of an error. For example, Hafner[57] has reviewed experimental volumes of mixing (called volume of formation in his paper) for 49 compounds of different categories, and found that many of them have volumes of mixing greater than our calculated maximum of 5% for γ(U,Zr) at $U_{eff}$ = 1 eV. Although the compounds summarized by Hafner are ordered phases, not solid solutions as is γ(U,Zr), these results suggest that our volume of mixing for γ(U,Zr) is not particularly abnormal.

Finally, in the third aspect of our analysis of the volume of γ(U,Zr), given the particularly large increase of volume from the noSOC to SOC case at $U_{eff}$ = 1.24, we consider explicitly the relative volume expansion due to SOC, which is shown in Fig. 1 c). Söderlind *et al.* argue that the volume expansion due to SOC in our DFT + *U* ($U_{eff}$=1.24 eV) calculations of γ(U,Zr) is too large because the expected range is about 1-2%[58,59] as demonstrated, for example, by their own DFT calculations of γ(U,Zr) using FPLMTO[51]. We agree with this statement. However, we disagree with Söderlind *et al.*'s assertion that this proves in general that "the combination of DFT + *U* and SOC" "results in unrealistically large"/"extreme" volume expansion. Here we demonstrate that this generalization is not correct with two sets of data, both provided in Fig. 1 c). The first set is volume expansions due to SOC at $U_{eff}$=1.24 eV for other known stable solid phases of U and U-Zr, which all can be calculated from the noSOC/SOC volume data we have previously provided in Table IV of Ref.[2]. The results clearly show no anomalous volume expansion due to SOC for all phases of U and U-Zr except γ(U,Zr). The second set of data in Fig. 1 (c) is volume expansion due to SOC for γ(U,Zr) at $U_{eff}$ = 1 eV (the single-structure optimized $U_{eff}$ for γ(U,Zr)). It is clear that they are also in the expected range. These results show that the large volume expansion due to SOC of γ(U,Zr) at $U_{eff}$=1.24 eV is not representative of our general results for other phases nor at other $U_{eff}$ values, and therefore does not suggest a fundamental problem with applying DFT + *U* for modeling U and U-Zr.

Overall, we reiterate that it is difficult to assess how well the volume of γ(U,Zr) is modeled by DFT and DFT + *U* due to a lack of sufficient and reliable experimental volume data, the mechanical instability of this phase, and the approximate estimation of finite temperature effects. However, our best attempts seem to suggest the following. DFT underestimates the volume of γ(U,Zr), especially the end member γU. At $U_{eff}$ = 1 eV, which is γ(U,Zr)'s single-structure optimized $U_{eff}$, DFT + *U* gives agreement with assessed experimental absolute volume data as good as or better than DFT. The predicted volume of mixing by DFT + *U* at $U_{eff}$ = 1 eV is larger compared to DFT, but it is not clear that this is incorrect given the constrained nature of the calculations, the limited experimental data and the necessity to extrapolate experimental data from over 1000 K to 0 K. Also the volume expansion due to SOC is within the expected 1-2% range for $U_{eff}$ = 1 eV. At the multi-structure optimized $U_{eff}$ = 1.24 eV for the overall U and U-Zr systems, especially when the calculations include SOC, DFT + *U* seem to give some unusual behavior and must be considered somewhat unreliable for γ(U,Zr). However, this only suggests an unphysical interaction of SOC with Hubbard *U* potential at $U_{eff}$ = 1.24 eV for γ(U,Zr), rather that a general failure of the DFT + *U* approach. Despite uncertainty for these findings, they are consistent with the idea that DFT + *U*, with proper $U_{eff}$, can yield improved results compared to DFT for the U-Zr system.

### 3.2.2. Bulk modulus (rebuttal to argument (b2))
We address in this section Söderlind *et al.*'s criticism (b2) that DFT + *U* gives worse bulk modulus than DFT for γU compared to the experimental value at 1100 K from Yoo *et al.*[17]. Although αU's bulk modulus is not mentioned by Söderlind *et al.*, we also examine it to complete the discussion we have started in Section 2.1.1.

*γU*

In this section we consider the bulk modulus for γU. Söderlind *et al.*[7] report that DFT + *U* at $U_{eff}$ = 2 eV underestimates the bulk modulus of γU considerably compared to the experimental value at 1100 K from Yoo *et al.*[17]. We obtain similar results to Söderlind *et al.*'s in our calculations at $U_{eff}$ = 2 eV. However, similar to volume, we again find that DFT + *U* can probably give improved bulk modulus for γU, again near γU's single-structure optimized $U_{eff}$ of 1 eV.

We begin by reviewing the experimental data. For γU, we only find one experimental reference Yoo *et al.* [17], which as mentioned in Section 2.1.1 is one of the earlier X-ray studies subject to nonhydrostatic stress. Unlike αU, the error associated with the nonhydrostatic media is difficult to estimate for γU due to limited data so we will estimate it based on similar studies of αU. The latest and perhaps most accurate value of αU's bulk modulus obtained from X-ray studies in quasihydrostatic media is 114.5 GPa at room temperature (298 K) [27]. In contrast, Yoo *et al.*[17] report a significantly larger value of 135.5 GPa at room temperature, which suggests that the error due to nonhydrostatic stress in their study[17] is about 135.5 − 114.5 = 21 GPa. Yoo *et al.*[17] report γU's bulk modulus at 1100 K to be 113.3 GPa. Because they use the same nonhydrostatic media when measuring both αU and γU, similar error to what we just calibrate for αU might be expected to occur in the γU measurements. Therefore, the true value for γU's bulk modulus at 1100 K is reasonably estimated to be 113.3 − 21 = 92.3 GPa. Before comparing to our *ab initio* results at 0 K, we need to extrapolate this experimental bulk modulus value measured at finite temperatures to zero temperature (0 K). There are accurate measurements of αU's bulk modulus at multiple temperatures between 298 and 923 K[31] that span the major part of its stable temperature range. Unfortunately, there is no similar data for γU on the evolution of bulk modulus with temperature as far as we are aware, and γU is not even stable below 1045 K[46]. Considering the challenges of obtaining the trend in γU's bulk modulus with temperature, we assume αU and γU have similar temperature dependence of bulk modulus and estimate that γU's bulk modulus should increase from 92.3 GPa at 1100 K to 109.5 GPa at 0 K (an increase of 17.5 GPa) based on αU's trend measured in Ref.[31]. We note that this effective bulk modulus for γU at 0 K is quite uncertain due to the necessity of extrapolating from high-temperature based on αU's trend and the fact that γU is not stable at lower temperatures. Normally we would not even attempt to compare 0K *ab initio* and high temperature experimental data in such detail for a phase that is both thermodynamically and mechanically unstable at 0 K. However, to address the criticisms of Söderlind *et al.*[7] about the calculated γU's bulk modulus, we will use this value as our current best-effort estimation.

Now we compare our *ab initio* bulk modulus to the estimated 0 K experimental values of 114.4 and 109.5 GPa for αU and γU, respectively, in Figure 3. This figure shows several important points that are common for both αU and γU: 1) DFT overestimates the bulk modulus substantially (by about 20 GPa or more); 2) DFT + *U* obtains smaller bulk modulus (the larger $U_{eff}$, the smaller the bulk modulus) and is in better agreement with experiment than DFT when $U_{eff}$ is in the range of 1-1.5 eV, in excellent agreement with the range of single-structure optimized $U_{eff}$'s we have determined in Ref.[2]; 3) DFT + *U* best reproduces the experimental bulk modulus of αU and γU near $U_{eff}$ = 1.5 eV and 1 eV, respectively, but considering the two phases together, DFT + *U* at $U_{eff}$ = 1.24 eV seems again to be near the multi-structure optimized $U_{eff}$ value, consistent with what we have obtained based on energetics in Ref.[2].

Let us focus on γU and make a more quantitative comparison. We find that DFT + *U* noSOC/SOC calculations at γU's single-structure optimized $U_{eff}$ of 1 eV obtain a bulk modulus of 107.6/103.3 GPa, which is only 1.9/6.2 GPa different from estimated 0 K experimental value of 109.5 GPa and should likely be within the error bars of the experimental data and the finite temperature extrapolation. At U and U-Zr's multi-structure optimized $U_{eff}$ of 1.24 eV, DFT + $U$ noSOC/SOC calculations give 97.1/90.9 GPa, respectively. The result is about 11/17% below the extrapolated 0 K experimental value. Even with these errors, the results are still comparable or better than those from DFT noSOC/SOC calculations, which in our own study with PAW obtain 135.1/128.5 GPa (23/17% above the extrapolated 0 K experimental value), while in Söderlind *et al.*[7]'s study give 134/128 GPa (23/17% above the extrapolated 0 K experimental value) with PAW and 137/150 GPa (23/37% above the extrapolated 0 K experimental value) with FPLAPW. In short, similar to volume, we find that γU's bulk modulus seem to be improved by DFT + $U$ at γU's single-structure optimized $U_{eff}$ of 1 eV when compared to DFT. The accuracy deteriorates at $U_{eff}$ = 1.24 eV, which is the multi-structure optimized $U_{eff}$ for the overall U and U-Zr systems, but is still better than DFT.

*αU*

Next we return our attention to αU. As described in Section 2.1.1, DFT calculations with SOC included using FPLMTO[9,18] or FPLAPW [32,33] give values that are larger than the 46 K experimental reference of 111.5 GPa from Fisher and McSkimin[29] by 21.5-36.5 GPa (19-33%) and neglecting SOC leads to an further increase in error to approximately 27-41%. As shown in Figure 3 a), our DFT results from PAW are similar, which are 143.3 and 141.0 GPa from noSOC and SOC calculations, larger than the 46 K reference by 31.8 and 29.5 GPa (28.5 and 26.5%), respectively. In comparison, DFT + $U$ obtains 136.7/127.1 and 132/124 GPa at $U_{eff}$ = 1.24 and 1.5 eV from our noSOC/SOC calculations respectively, which are lower than DFT and closer to the experimental values. These results demonstrate that, like enthalpy and volume, DFT + $U$ can also improve the accuracy for modeling the bulk modulus of αU. We expect DFT + $U$ to be also of value for modeling the closely related property of elastic constants that we have discussed above in Section 2.1.2.

### 3.2.3. Enthalpy (rebuttal to argument (b3))

*γ(U,Zr)*

Söderlind *et al.*[7] argue that (b3) DFT + $U$ ($U_{eff}$=1.24 eV) with SOC calculates the enthalpy of mixing of γ(U,Zr) to be negative for a large range of composition and that this result is inconsistent with the known miscibility gap in γ(U,Zr). However, this argument is not really evidence against the DFT + $U$ approach for the following three reasons. Firstly, only a full miscibility gap disproves negative enthalpy of mixing—a partial miscibility gap can well be produced by an enthalpy that is negative in some region. As Söderlind *et al.* themselves suggest, our DFT + $U$ enthalpy can still give a miscibility gap approximately between 0-70 at.%Zr. Experimentally, the miscibility gap of γ(U,Zr) shows in the phase diagram approximately between 10 and 50 at.%Zr[60], and thus is still within the range consistent with our DFT + $U$ ($U_{eff}$=1.24 eV) enthalpy. It is possible γ(U,Zr) indeed has a full miscibility gap but is covered in part by competing phases, as the CALPHAD models[1,61,62] suggest, in which case the DFT + $U$ ($U_{eff}$=1.24 eV) predictions of negative formation enthalpy would then be incorrect. However, without any direct data this comparison is not very strong evidence for or against the DFT + $U$ approach. Secondly, our DFT + $U$ ($U_{eff}$=1.24 eV) results are within our expected error bars of being positive. The minimum in our DFT + $U$ ($U_{eff}$=1.24 eV) enthalpy is about -0.01 eV/atom (1 kJ/mole). We have shown in FIG. 2 and 3 of Ref.[2] that even at $U_{eff}$=1.24 eV our DFT + $U$ calculated enthalpies have typical errors compared to experimental and/or CAPHAD references of 0.02-0.04 eV/atom, which means that for γ(U,Zr) obtaining negative 0.01 eV/atom is potentially consistent with a positive mixing value. Thirdly and most importantly, as we have been suggesting when discussing volume and bulk modulus in previous sections, the single-structure optimized $U_{eff}$ of γ(U,Zr) is near 1 eV and not equal to the multi-structure optimized $U_{eff}$ of 1.24 eV. At $U_{eff}$ = 1 eV, the enthalpy of mixing of γ(U,Zr) is positive at all compositions. In fact, current evidence tends to support that DFT + $U$ probably describes γ(U,Zr) closer to its true thermodynamics than DFT— DFT + $U$ gives enthalpy with the same asymmetry vs. composition seen in the result from an electromotive force experiment[63] at 1073 K while the enthalpies from DFT calculations of Landa *et al.*[51] and us[2] and the two early CALPHAD models[61,62] seem to be quite symmetrical; also, a very recent independent study[64] employing modified embedded atom method potential finds their result to be closer in nature to those from our CALPHAD[1] and DFT + $U$[2] than DFT from Landa *et al.*[51] and us[2]. In short, at U and U-Zr's overall multi-structure optimized $U_{eff}$ of 1.24 eV, DFT + $U$ can still give enthalpy of mixing for γ(U,Zr) that is consistent with the experimental miscibility gap of γ(U,Zr) when the compositional range of the miscibility gap and uncertainty of the calculations are considered; even better, at γ(U,Zr)'s single-structure $U_{eff}$ of 1 eV DFT + $U$ shows promise of improving the enthalpy of γ(U,Zr) vs. DFT.

### 3.2.4. Magnetic moments (rebuttal to argument (b4))

Söderlind *et al.*[7] argue that b4) DFT + $U$ "predict significant spin and orbital magnetic moments for most phases of uranium and all γ-U-Zr alloys", which they believe suffice to keep DFT + $U$ from being applied on modeling U and U-Zr because magnetic moments are not believed to occur in nature for U metal, although their existence (or non-existence) is uncertain for U-Zr alloy due to lack of experimental data. While we agree with Söderlind *et al.*[7]'s observation that DFT + $U$ may give unphysical magnetic moments for U and U-Zr when $U_{eff}$'s are large enough, we disagree with Söderlind *et al.*'s discussion[7] of the magnetic moments issue on three aspects.

Our first disagreement is that it is hasty generalization to claim DFT + $U$ in general leads to unphysical magnetic solutions. On the one hand, we have shown[2] that at $U_{eff}$ = 0.75 or even close to 1 eV, DFT + $U$ still predicts mostly vanishing spin and orbital moments for all the three solid phases of U metal. On the other hand, we have already explained in Ref.[2] (p 13) that for γ(U,Zr) "the presence of these moments may be an artifact of the constrained relaxations". In fact, we have shown[2] that even DFT still predicts finite spin and orbital moments for γ(U,Zr) at four out of the five compositions studied. Even for α(Zr) alloy at 93.8 at.%Zr without the constraint, DFT also predicts small but finite spin and orbital moments. It is thus unfair to attack only DFT + $U$ as the culprit for promoting magnetic moments.

Our second disagreement is on abandoning DFT + $U$ completely due to magnetic moments. Admittedly, Söderlind *et al.* have rightfully pointed out that significant unphysical moments are predicted with DFT + $U$ for βU and γU at $U_{eff}$=1.24 eV, which is the multi-structure optimized $U_{eff}$ we have suggested, and more generally within the range of 1-1.5 eV for single-structure optimized $U_{eff}$'s of U and U-Zr. We fully agree that these magnetic solutions suggest that the present DFT + $U$ approach is not completely satisfactory. As we have noted in Ref.[2], such an imperfection is not surprising given that DFT + $U$ is equivalent to the DFT + DMFT formalism with the DMFT impurity problem solved within the Hartree-Fock approximation[44]. Hartree-Fock can only incorporate correlations via real, static self-energies which amount to splitting in the spin and orbital sectors. The Hartree-Fock self-energy therefore often exaggerates polarization as compared to the complex, frequency dependent self-energy that can be obtained exactly using quantum Monte-Carlo[65]. However, in light of the present DFT + $U$ results, given that energetics, volume, bulk modulus can be generally improved, we believe it would be misguided to abandon the approach just because of the emergence of spurious magnetic moments. This situation is in agreement with the case of an early study by Söderlind and Sadigh[66] in which the energetics of the six known allotropes of Pu metal are concluded to be well reproduced by their DFT plus orbital polarization (DFT + OP) calculations despite the fact that significant static local magnetic moments are obtained in the calculations (> 2.0 μB/atom, see FIG. 3 of Ref.[22]) and not seen in the experiments[67]. Furthermore, we hope that future work will address this problem of spurious moments in DFT + $U$ by using DFT + DMFT, with which the static moments will perhaps not be predicted, while the same or better quantitative improvements are expected for the other properties. As an example, for δPu, one of the allotropes of Pu, DFT + $U$ also gives in a magnetic solution in Ref.[68], while DFT + DMFT using quantum Monte-Carlo produces a mildly correlated Fermi liquid[69]. Another factor to consider in addressing the undesirable moments besides the solution to the DMFT impurity problem is the double-counting correction scheme, which could also be responsible for pushing the system into an excessively correlated regime. Both our study[2] and that of Söderlind *et al.*[7] use the so-called fully localized limit (FLL) double-counting, but there are other choices that could be more appropriate. In particular, the so-called around the mean field (AMF) scheme has been shown to give magnetic polarization a much larger energy penalty than the FLL[70]. Returning to the example of δPu, the FLL based DFT + $U$ also results in a magnetic solution in Ref. [68], but AMF based DFT + $U$ in Ref. [71] yields a nonmagnetic ground state, in agreement with experiment results. The AMF based DFT + $U$ in Ref. [71] also reproduces other equilibrium properties well, and significantly better than the conventional DFT. Further study is needed to assess to what extent different double counting correction schemes in DFT + $U$ might reduce or remove the moments in the range of physical $U_{eff}$ values.

Our last disagreement is on Söderlind *et al.*'s interpretation that we have claimed that "anti-parallel spin and orbital contributions nearly cancel" are consistent with "non-magnetic state of uranium metal" in Ref.[2]. We realize that this interpretation is possibly caused by our statement that our DFT + $U$ calculations "correctly predict the magnetic structure" (p.13 of Ref.[2]). However, the context of this sentence indicates that by "magnetic structure" we mean that moments, if exist are quite random in terms of magnitude and direction not in "long-range ferromagnetic or antiferromagnetic ordering". Moreover, our discussion in Ref.[2] is restricted to magnetic moments and structure, not magnetic state, and our assertion that cancelling moments yield no net *moment* is not equivalent to mean non-magnetic *state*. In fact, we specifically state that magnetic moments "are difficult to compare to experiment" (p.14 of Ref.[2]). However, we realize that the two concepts are indeed very close, and our discussion in Ref.[2] is unclear enough that people may interpret our statements as meaning vanishing local total moments to be equivalent with non-magnetic state. Therefore, we appreciate Söderlind *et al.* bringing up this issue and take this chance to reiterate that a state with finite but canceling spin and orbital magnetic moments is not a nonmagnetic state.

### 3.2.5. Optimal $U_{eff}$ for γ(U,Zr) (rebuttal to argument (b5))

The last remaining unaddressed criticism of Söderlind *et al.* is (b5) that DFT + $U$ needs to use greatly different $U_{eff}$'s for the enthalpy and volume of γ(U,Zr). This assertion is not true. As we discussed extensively above, $U_{eff}$ of 1 eV is close to the optimal value for both enthalpy and volume of γ(U,Zr). For the related γU, the corresponding value is also 1 eV for enthalpy, volume and bulk modulus. As defined in Sec. 3.1, 1 eV is γU and γ(U,Zr)'s single-structure optimized $U_{eff}$. These results show that for a single system like γ(U,Zr), a fairly consistent empirical $U_{eff}$ exists for different properties and no excessive parameter-fitting procedure is necessary.

# 4. Summary

We believe that this work helps clarify a number of confusions and provides additional data of value for assessing the applicability of DFT + $U$ to U metal and U-Zr alloy. We have made the following major points in this article:

1) The long-held belief that DFT based on GGA, especially solved using all-electron methods, models bulk modulus and elastic constants of αU well is questionable because of issues existing in the experimental references commonly used to benchmark theory in previous studies. Old but often neglected experimental results, as well as very recent experimental results suggest that DFT even when solved with the all-electron method FPLMTO[9] still gives a relative error of 27-44% and 19-33% for bulk modulus in noSOC and SOC calculations, respectively, and approximately a root mean square error of 49% for elastic constants in SOC calculations, both being significantly larger than found in other comparable systems (e.g., in most transition metals).

2) The difference in the volume of αU between our PAW and Söderlind's FPLMTO origin from sources including different GGA functionals, structural relaxations, and implementations of SOC in addition to the PAW approximation. Our quantitative comparison that also considers the value from the FPLMTO study by Le Bihan *et al.*[18] suggests that our PAW calculations for U and U-Zr do not suffer from any apparent "deficiency" as claimed by Söderlind *et al.* Based on both 1) and 2), we maintain our original conclusion that accuracy of DFT for modeling U and U-Zr is not as good as that for other systems (e.g., transition metals) and have room for improvement.

3) One major issue of Söderlind *et al.*'s criticisms of DFT + $U$ is that they have focused primarily on the high temperature BCC phases γU and γ(U,Zr), which are thermodynamically and mechanically unstable at the modeling temperature of 0 K, while neglected results for most of the other phases that are actually primary evidence of our arguments for DFT + $U$. In particular, Söderlind *et al.* neglect comparisons to enthalpies for the stable solid phases of U and U-Zr other than γ(U,Zr). As a result, Söderlind *et al.*'s major claim that DFT + $U$ gives worse modeling results than DFT is in general not supported when one considers all the U and U-Zr phases and when one considers the lack of reliable low temperature references for γU and γ(U,Zr).

4) Another main issue of Söderlind *et al.*'s criticisms of DFT + $U$ is an overgeneralization of results at selected $U_{\text{eff}}$ values to argue that DFT + $U$ in general suffers from issues that occur at specific $U_{\text{eff}}$ values or ranges of values. In their Letter[7], they have used only $U_{\text{eff}}$ = 2 eV in their DFT + $U$ calculations of γU. Although this value is suggested from theoretical estimation of Hubbard $U$ for U metal, it is too large because DFT + $U$ is based on Hartree-Fock approximation that is expected to overestimate the effects of the Hubbard $U$[44]. Use of this large value leads to many of the undesired results which lead Söderlind *et al.* to find fault with DFT + $U$. Similarly, in their Comment[8] Söderlind *et al.* have focused on the "multi-structure optimized $U_{\text{eff}}$" of 1.24 eV, which is expected to give a minimal *average* error when considering many phases in the U and U-Zr systems together. Söderlind *et al.* have rightfully pointed out that if we model γU and γ(U,Zr) with U and U-Zr's multi-structure optimized $U_{\text{eff}}$=1.24 eV, DFT + $U$ probably gives too large absolute volume and volume expansion due to SOC on the U-rich end. However, we show that at γU and γ(U,Zr)'s single-structure optimized $U_{\text{eff}}$=1 eV, which is expected to give optimal results specifically for γU and γ(U,Zr), DFT + $U$ results for volume, bulk modulus and enthalpy of γU and γ(U,Zr) do not suffer from the issues which Söderlind *et al.* criticize. In fact, DFT + $U$ with $U_{\text{eff}}$=1 eV generally gives improved results compared to DFT for γU and γ(U,Zr). That said, we reiterate our belief that validation of *ab initio* approaches using the high temperature γU and γ(U,Zr) BCC phases are very unreliable because of the instability at low temperature and the lack of low temperature experimental data for these phases. Overall we believe that the errors for γU and γ(U,Zr) at $U_{\text{eff}}$ = 1.24 eV are not so severe that the whole DFT + $U$ approach should be abandoned for the general systems of U and U-Zr. Instead, they illustrate that using the single multi-structure optimized $U_{\text{eff}}$ = 1.24 eV across different structures of U and U-Zr may not always be perfect for each individual structure. If available, single-structure optimized $U_{\text{eff}}$ should be used for modeling individual structures. It not, the multi-structure optimized $U_{\text{eff}}$ is a good start but additional explorations may be helpful to avoid errors like those we see for γU and γ(U,Zr) at $U_{\text{eff}}$ = 1.24 eV.

6) Some of the key claims that Söderlind *et al.* use to criticize DFT + $U$ are simply not true. These include their claims that $U_{\text{eff}}$ = 1.24 eV gives unprecedented large positive deviation from linear dependence of composition (FIG 1 of their Comment[8]) and that the partially negative enthalpy of mixing for γ(U,Zr) is inconsistent with the existence of miscibility gap in the experimental phase diagram (FIG 3 of their Comment[8]).

7) We agree with Söderlind *et al*. that the emergence of magnetic moments in select phases of U metal and U-Zr alloy at $U_{\text{eff}}$ between 1-1.5 eV is probably unphysical. However, we do not share Söderlind *et al*.'s belief that this implies we should avoid modeling U and U-Zr with DFT + $U$ completely. We have shown that despite unphysical moments DFT + $U$ appears to provide many improved ground state properties, like enthalpy, volume, and bulk modulus, when compared to DFT. Furthermore, these artificial moments are not necessarily inherent to the Hubbard $U$ modification and can potentially be avoided if one uses alternative double counting terms in DFT + $U$, or go beyond the Hartree-Fock approximation and use DFT + DMFT, whereby the DMFT impurity problem is solved exactly, e.g., via quantum Monte-Carlo.

8) Finally, Söderlind *et al*. have also incorrectly represented our opinions on three issues. First we have not argued that U and U-Zr is strongly correlated but instead we think electron correlation strength in them is small to intermediate. Second, we have not argued that one must include SOC when modeling U and U-Zr but instead that including SOC can improve the calculated properties but the improvement is small. Third, we do not think cancellation of spin and magnetic moments is a nonmagnetic state. However, in all these cases we realize that that our previous statements may have been unclear and that some readers might have come to the conclusions of Söderlind *et al*. Therefore, we have used this article to clarify our opinions on these issues.

# Acknowledgement

This research was funded by the US Department of Energy Office of Nuclear Energy's Nuclear Energy University Programs under Contract No. 00088978. We acknowledge computing time from Idaho National Laboratory's Center for Advanced Modeling and Simulation. This work also used the Extreme Science and Engineering Discovery Environment (XSEDE), which is supported by National Science Foundation grant number ACI-1053575. We are grateful to P. Söderlind for providing us the original numerical data published in Fig. 9 of Ref.[51] and to A. Landa for suggestions on experimental volume data for γ(U,Zr).

# Table

**Table 1.** Elastic constants of αU from DFT theory and ultrasound experiment.

| Elastic constants | Theory (GPa) | Expt. (GPa) | | | | Relative error (RE)[f] | | | |
|---|---|---|---|---|---|---|---|---|---|
| | 0 K[a] | 298 K[b] | 46 K[c] | 4.2 K[d] | 0 K[e] | 298 K | 46 K | 4.2 K | 0 K |
| $c_{11}$ | 300 | 214.8 | 161.0 | 114.3 | 215.0 | 40% | 86% | 162% | 40% |
| $c_{22}$ | 220 | 198.6 | 209.0 | 211.1 | 199.8 | 11% | 5% | 4% | 10% |
| $c_{33}$ | 320 | 267.1 | 287.5 | 286.0 | 269.6 | 20% | 11% | 12% | 19% |
| $c_{44}$ | 150 | 124.4 | 141.1 | 139.6 | 126.2 | 21% | 6% | 7% | 19% |
| $c_{55}$ | 93 | 73.4 | 90.1 | 82.0 | 75.3 | 27% | 3% | 13% | 24% |
| $c_{66}$ | 120 | 74.3 | 85.1 | 89.2 | 75.4 | 62% | 41% | 35% | 59% |
| $c_{12}$ | 50 | 46.5 | 29.4 | 28.6 | 45.8 | 8% | 70% | 75% | 9% |
| $c_{13}$ | 5 | 21.8 | 32.0 | 34.7 | 21.6 | -77% | -84% | -86% | -77% |
| $c_{23}$ | 110 | 107.6 | 111.8 | 112.9 | 107.8 | 2% | -2% | -3% | 2% |
| Mean absolute error (MAE)[f] | | | | | | 30% | 34% | 44% | 29% |
| Root mean square error (RMSE)[f] | | | | | | 38% | 49% | 67% | 37% |

[a]Söderlind's DFT calculation using FPLMTO with SOC included[9].
[b]Fisher and McSkimin[22].
[c]Measured by Fisher and McSkimin[29] and tabulated by Fisher[31].
[d]Fisher and Dever[37].
[e]Extrapolated to 0 K by McSkimin and Fisher from polynomial fitting of the values they measure between 78 and 298 K[36].

[f]RE, MAE, RMSE are defined respectively as $\mathrm{RE}(c_i) = (c_i^{\mathrm{Theory}} - c_i^{\mathrm{Expt.}})/c_i^{\mathrm{Expt.}}$, $\mathrm{MAE} = \sum_{i=1}^{n} |\mathrm{RE}(c_i)|/n$,

$\mathrm{RMSE} = \sqrt{\sum_{i=1}^{n} \mathrm{RE}(c_i)^2 / n}$, where $i \in \{11, 22, \cdots 23\}$ and $n=6$.

**Table 2**. Volume of γ(U,Zr) in unit of Å³/atom from a) experiment, b) theory in this work, and b) theory in the literature. γU is one end member of γ(U,Zr) with 0 at.%Zr and βZr is the other end member with 100 at.%Zr. SOC and noSOC denote calculations with and without spin orbit coupling included, respectively.

a) Experiment. The three experiments Lawson *et al.*[46], Akabori *et al.*[49], and Heiming *et al.*[47] directly measure volumes at high temperatures where γ(U,Zr) is stable, and both the original finite temperature data and the estimated 0 K values are given here (see supplementary materials of Ref.[2] for details of the estimation). The other two experiments, Huber and Ansari[48] and Basak *et al.*[50] measure quenched samples at room temperature and the original data are tabulated directly here.

| Mole Fraction of Zr | Lawson *et al.* | Huber and Ansari (room *T*) | Akabori *et al.* | Basak *et al.* (room *T*) | Heiming *et al.* |
|---|---|---|---|---|---|
| 0 | 22.05/21.46 (1060/0 K) | | | | |
| 0.25 | | 22.37 | | | |
| 0.3 | | 22.24 | | | |
| 0.4 | | 22.05 | | | |
| 0.5 | | 22.29 | | | |
| 0.5 | | 22.29 | | | |
| 0.6 | | 22.52 | | | |
| 0.7 | | 22.62 | | | |
| 0.707 | | | 23.11/22.78 (925/0 K) | | |
| 0.723 | | | | 22.87 | |
| 0.749 | | 22.75 | | | |
| 0.8 | | 22.8 | | | |
| 1 | | | | | 23.7/23.1 (1253/0 K) |

b) Theory in this work. All are calculated using PAW and have been reported in Table IV and FIG. 6 of Ref.[2] except the DFT + *U* (1eV) SOC value (explained in note 1 below).

| Mole Fraction of Zr | DFT (0 K) | | DFT + *U* (1eV) (0 K) | | DFT + *U* (1.24eV) (0 K) | |
|---|---|---|---|---|---|---|
| | noSOC | SOC | noSOC | SOC | noSOC | SOC |
| 0 | 20.13 | 20.17 | 20.98 | 21.18[1] | 21.28 | 22.77 |
| 0.0625 | 20.36 | 20.41 | 21.38 | 21.63 | 21.96 | 22.79 |
| 0.25 | 21.10 | 21.18 | 22.23 | 22.60 | 22.62 | 23.20 |
| 0.5 | 21.97 | 22.06 | 22.75 | 22.99 | 23.02 | 23.33 |
| 0.75 | 22.43 | 22.60 | 22.89 | 23.00 | 23.04 | 23.23 |
| 0.9375 | 22.88 | 22.86 | 22.94 | 22.94 | 23.01 | 22.97 |
| 1[2] | 22.91 | 22.91 | | | | |

Note 1: We have reported a solution with volume of 21.51 Å³/atom in Fig. 6 of Ref.[2], but more recently find another solution to be 0.001 eV/atom lower in energy, the volume of which (21.18 Å³/atom) is then used here as the ground state solution.
Note 2: DFT + *U* is not applied on Zr.
c) Theory in the literature. The two references are Söderlind *et al.*[7] and Landa *et al.*[51].

| Mole Fraction of Zr | DFT-KKRASA (300 K)[3] | DFT-FPLMTO (0 K)[4] | DFT-FPLAPW (0 K)[5] | |
|---|---|---|---|---|
| | noSOC | noSOC | noSOC | SOC |
| 0 | 21.4 | | 20.3 | 20.6 |
| 0.1 | 21.7 | | | |
| 0.2 | 22.0 | | | |
| 0.3 | 22.3 | | | |
| 0.4 | 22.7 | | | |
| 0.5 | 23.0 | | | |
| 0.6 | 23.2 | | | |
| 0.7 | 23.4 | | | |
| 0.8 | 23.6 | | | |
| 0.9 | 23.7 | | | |
| 1 | 23.8 | 22.98 | | |

Note 3: From Fig.1 (a) of Landa *et al*.[51].
Note 4: From Fig. 9 of Landa *et al*.[51]. The original numerical values were provided to us by P. Söderlind in a private communication on November 20$^{th}$, 2013.
Note 5: From Table 1 of Söderlind *et al*.[7].

# Figures

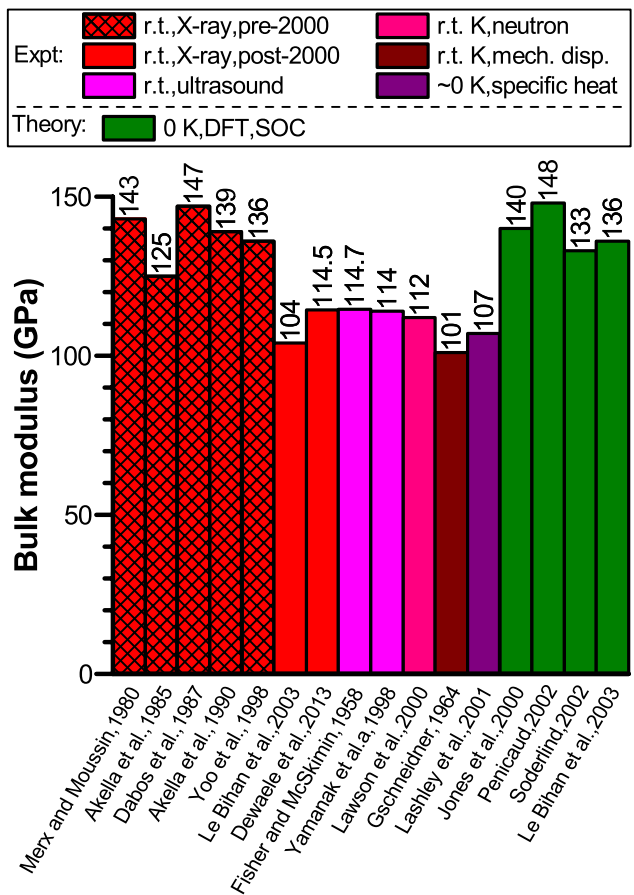

**Figure 1.** Bulk modulus of αU measured using various experiment techniques[17-27] and from DFT calculated using all-electron methods with SOC included[9,18,32,33]. The experimental data reviewed in Table I of Ref.[18] are quoted in addition to the very recent X-Ray result of Dewaele *et al*.[27]. The value of Fisher and McSkimin[22] has been incorrectly cited in Ref.[18] but corrected here. r.t. in the legend denotes room temperature.

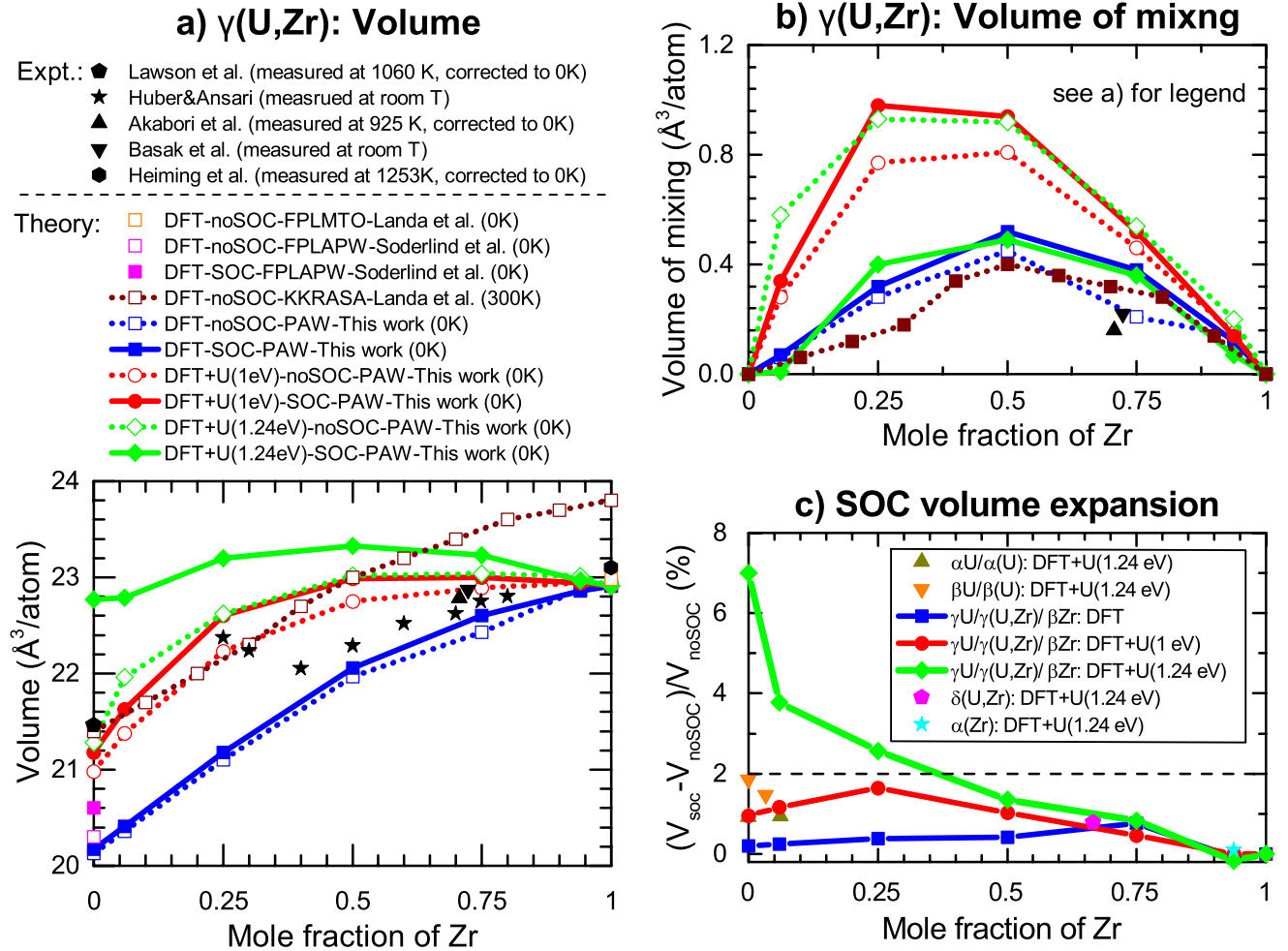

**Figure 2**. γ(U,Zr)'s a) volume, b) volume of mixing, and c) volume expansion due to spin-orbit coupling (SOC). γU is one end member of γ(U,Zr) with 0 at.%Zr and βZr is the other end member with 100 at.%Zr. SOC and noSOC denote calculations with and without SOC included, respectively. Estimated 0 K values are plotted here for the three experimental studies of Lawson *et al.* [46], Akabori *et al.*[49], and Heiming *et al.*[47] that directly measure volumes at high temperatures where γ(U,Zr) is stable, while the original values are plotted for the other two experiments, Huber and Ansari[48] and Basak *et al.*[50] that measure quenched samples at room temperature. Note Huber and Ansari[48]'s curve shows a convex curvature opposite to what is expected for a phase separating phase and should be considered less trustworthy (see main texts). Volume of mixing for γ(U,Zr) is defined as $V_{\gamma(U,Zr)}^{mix} = V_{\gamma(U,Zr)} - (1-x)V_{\gamma U} - xV_{\beta Zr}$ where $x$ is Zr mole fraction. Experimental volume of mixing for Akabori *et al.*[49] and Basak *et al.* [50] is calculated by referencing to Lawson *et al.* for γU[46] and Heiming *et al.* for βZr[47], while Huber and Ansari[48] is neglected due to its unrealistic convex curvature (see text for details on these choices). Volume expansion due to SOC is calculated as $(V_{SOC} - V_{noSOC})/V_{noSOC}$. Panel c) includes not only γ(U,Zr) but also all phases that we have tabulated in Table IV of Ref.[2].

.

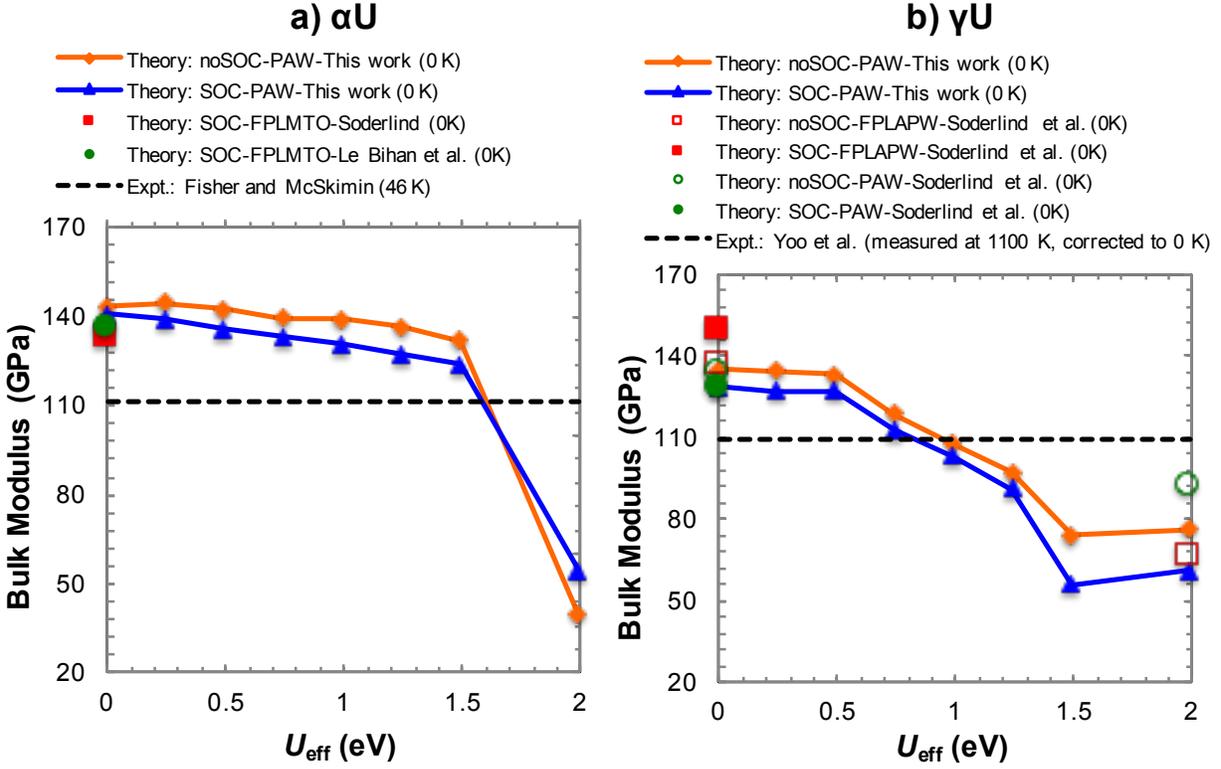

**Figure 3**. Bulk modulus of a) αU and b) γU as a function of $U_{\text{eff}}$. SOC and noSOC in the legend denote calculations with and without spin-orbit coupling (SOC) included, respectively. For αU, the two DFT results are from FPLMTO calculations of Söderlind[9] and Le Bihan *et al.*[18] while the experimental value measured at 46 K is from Fisher and McSkimin[29]. For γU, the DFT results are from Söderlind *et al.* [7] and the nonhydrostatic stress error corrected experiment value extrapolated to 0 K (see main texts) is from Yoo *et al.* [17].